\documentclass[12pt]{article}
\usepackage{latexsym}
\usepackage{amsmath,,calrsfs}
\usepackage{amsfonts}
\usepackage{amssymb}
\usepackage{amscd}
\usepackage{bbm}
\usepackage{fancybox}
\usepackage{cite}
\usepackage{amsmath,amsfonts,amsbsy}
\usepackage{pstricks,pst-node}
\usepackage[small,bf,hang]{caption2}
\usepackage{graphicx}
\usepackage{epsfig}
\usepackage{psfrag}
\usepackage{comment}

\usepackage{float}

\psset{unit=1.3cm,linewidth=.5pt,radius=.2}  

\usepackage{multirow}                     
\usepackage{float}                          
\usepackage{lscape}                         
\usepackage{bm}

\newcommand{\bb}{\bar\beta}

\newcommand{\beq}{\begin{equation}}
\newcommand{\eeq}{\end{equation}}
\newcommand{\bi}{\begin{itemize}}
\newcommand{\ei}{\end{itemize}}
\newcommand{\bt}{\begin{tabular}}
\newcommand{\et}{\end{tabular}}
\newcommand{\bc}{\begin{center}}
\newcommand{\ec}{\end{center}}

\newcommand{\be}{\begin{equation}}
\newcommand{\ee}{\end{equation}}
\newcommand{\bea}{\begin{eqnarray}}
\newcommand{\eea}{\end{eqnarray}}
\newcommand{\ba}{\begin{array}}
\newcommand{\ea}{\end{array}}

\def\bbox{{\,\lower0.9pt\vbox{\hrule \hbox{\vrule height 0.2 cm
\hskip 0.2 cm \vrule height 0.2 cm}\hrule}\,}}
\newcommand{\dsl}{\pa \kern-0.5em /}

\font\mybb=msbm10 at 12pt
\def\bb#1{\hbox{\mybb#1}}
\def\bZ {\bb{Z}}
\def\bR {\bb{R}}
\def\bO {\bb{O}}

\def\bN {\bb{N}}
\def\bM {\bb{M}}
\def\bL {\bb{L}}
\def\bG {\bb{G}}
\def\bH {\bb{H}}
\def\bC {\bb{C}}
\def\bB {\bb{B}}
\def\bA {\bb{A}}
\def\bX {\bb{X}}
\def\bY {\bb{Y}}
\def\bP {\bb{P}}
\def\bJ {\bb{J}}
\def\bK {\bb{K}}
\def\bU {\bb{U}}
\def\bW {\bb{W}}
\def\bV {\bb{V}}
\def\bS {\bb{S}}
\def\bJ {\bb{J}}
\def\bM {\bb{M}}

\def\bI{\bb{I}}

\def\bfs{\mbox{\boldmath $\sigma$}}

\def\bfS{\mbox{\boldmath $\Sigma$}}
\def\bfpsi{\mbox{\boldmath $\psi$}}

\def\tr{{\rm tr}}

\def\bfs{\mbox{\boldmath $\sigma$}}

\def\tr{{\rm tr}}

\newcommand{\Pslash}{P\hskip -.24truecm  / }

\makeatletter \@addtoreset{equation}{section} \makeatother

\def\slashchar#1{\setbox0=\hbox{$#1$}           
   \dimen0=\wd0                                 
   \setbox1=\hbox{/} \dimen1=\wd1               
   \ifdim\dimen0>\dimen1                        
      \rlap{\hbox to \dimen0{\hfil/\hfil}}      
      #1                                        
   \else                                        
      \rlap{\hbox to \dimen1{\hfil$#1$\hfil}}   
      /                                         
   \fi}

\begin{document}

\begin{titlepage}
\begin{center}

\hfill  DAMTP-2016-14


\vskip 1.5cm

{\large \bf  Pauli-Lubanski, Supertwistors, and the Superspinning Particle}

\vskip 2cm

{\bf Alex S. Arvanitakis${}^{1,2}$, Luca Mezincescu${}^3$,  and 
Paul K.~Townsend${}^1$} \\

\vskip 15pt

{\em $^1$ \hskip -.1truecm
\em  Department of Applied Mathematics and Theoretical Physics,\\ Centre for Mathematical Sciences, University of Cambridge,\\
Wilberforce Road, Cambridge, CB3 0WA, U.K.\vskip 5pt }

{email: {\tt A.S.Arvanitakis@damtp.cam.ac.uk, P.K.Townsend@damtp.cam.ac.uk}} \\

\vskip .4truecm

{\em $^2$ \hskip -.1truecm
\em Department of Nuclear and Particle Physics,\\
Faculty of Physics, National and Kapodistrian University of Athens, \\
Athens 15784, Greece}

\vskip .4truecm

{\em $^3$ \hskip -.1truecm
\em Department of Physics
University of Miami, \\ P.O. Box 248046, Coral Gables, FL 33124, USA\vskip 5pt }

{email: {\tt Mezincescu@physics.miami.edu}} \\

\end{center}

\vskip 0.5cm

\begin{center} {\bf ABSTRACT}\\[3ex]
\end{center}

We present a novel construction of the  super-Pauli-Lubanski pseudo-vector for  4D supersymmetry and show how it arises naturally from the spin-shell constraints
in the supertwistor formulation of superparticle dynamics. We illustrate this result in the context of a simple classical  action for a ``superspinning particle'' of superspin 1/2.
We then use  an $Sl(2;\bK)$-spinor formalism for $\bK=\bR,\bC,\bH$ to unify our 4D results with previous results for 3D and 6D.

\end{titlepage}

\newpage
\setcounter{page}{1} 
\tableofcontents


\section{Introduction}

The spin of an elementary particle of non-zero mass is determined (e.g. in Wigner's classification of unitary irreps of the Poincar\'e group \cite{Wigner:1939cj})  by a choice of irreducible 
representation of $SU(2)$, the double cover  of the rotation group, which is itself determined by the choice of a non-negative integer or half-integer $s$. In the context of relativistic particle mechanics, which is our focus here, the incorporation of spin in a manifestly Lorentz covariant way (e.g. in Souriau's  classification of classical ``elementary systems'' \cite{Sou}) involves the Pauli-Lubanski (PL) polarization pseudo-vector 
$W$. For any representation of the Poincar\'e group, spanned by the generators of Minkowski spacetime translations ($P$) and  Lorentz ``rotations'' ($J$), this is defined as 
\begin{equation}\label{PLvec}
W^m = \frac{1}{2} \varepsilon^{mnpq} P_nJ_{pq}\, . 
\end{equation}
For a quantum system the product is the matrix product in the chosen representation but for many purposes it is sufficient to consider a Poisson bracket realization of the Poincar\'e 
algebra in terms of classical Noether charges, in which case the product is multiplication of functions on phase space. The Poincar\'e Casimirs are then the scalar functions $P^2$ and 
$W^2$, and for a particle of mass $m$ and spin $s$ we have $P^2=-m^2$ and $W^2= m^2s^2$. These Casimirs are zero for zero mass, in which case $W_m=hP_m$ for helicity $h$. 

It is convenient to replace the pseudo-vector $W^m$ by the 3-form 
\begin{equation}\label{PL3form}
W_{mnp} = P_{[m}J_{np]}\, , 
\end{equation}
because this has the advantage of being dimension independent: there is a PL 3-form in every dimension $d\ge3$. In general there are PL  $(2n+1)$-forms for $2n\le d-1$. For example, for 
a  Minkowski spacetime of dimension 5 or 6 (we abbreviate this to 5D,\, 6D etc) one also needs to consider the PL 5-form 
\begin{equation}
\Upsilon_{mnpqr} =  P_{[m}J_{np}J_{qr]}\, .
\end{equation}

One purpose of this paper is to provide constructions of the super-Pauli-Lubanski (SPL) tensors that have the same relevance to the classification of elementary superparticles as PL tensors have to the classification of elementary particles.  One might suppose
that this is a straightforward exercise in the conversion of a PL tensor into a super-PL tensor by the addition of terms that promote  translation invariance to super-translation invariance; however, 
 this is not so simply achieved. 

Consider the case of minimal  (${\cal N}=1$) 4D supersymmetry,  for which there is just one 4-component Majorana-spinor supersymmetry charge $Q$. If we assume a Poisson bracket realization of the super-Poincar\'e algebra then the components of $Q$ are {\it anticommuting} functions on the phase superspace of some super-Poincar\'e
invariant superparticle mechanics model.  We  might try to write down  a generalization of the PL  3-form that  is super-translation invariant, i.e. one that has zero Poisson brackets with the generators $\{P,Q\}$.  However, if we assume that it is polynomial in super-Poincar\'e generators with purely  numerical (i.e. dimensionless) coefficients  then all candidates have the  form\footnote{Here, $\bar Q= Q^TC$ for charge conjugation matrix $C$, and we recall that the matrices $C\Gamma_{mnp}$  are antisymmetric
in four spacetime dimensions.}
\begin{equation}\label{Wa}
W_{mnp}(a) = J_{[mn}P_{p]} -\frac{ia}{24} \bar Q \Gamma_{mnp}Q\, , 
\end{equation}
for some number $a$; this follows from a rescaling invariance of the  super-Poincar\'e algebra with scaling weights $[J]=0$,  $[P_m]=1$ and $[Q] = \frac{1}{2}$. 
The problem with this formula is that $W(a)$ is not supertranslation invariant for any value of $a$. 

This is a well-known problem. One standard resolution of it due to Salam and Strathdee \cite{Salam:1974yz} (see also \cite{Sokatchev:1975gg}, and \cite{Buchbinder:1998qv} for a detailed exposition) is to consider  the 2-form
$P^pW_{mnp}(2)$. This {\it is} supertranslation invariant  in our conventions (to be spelt out later) and  its norm squared is, in the quantum theory,  a super-Poincar\'e Casimir proportional to the quadratic Casimir $C_2$ of $SU(2)$. In fact, 
\begin{equation}\label{standard}
9\left[P^p W_{nmp}(2)\right] \left[P_qW^{mnq}(2)\right] = 2m^4 C_2 \, . 
\end{equation}
In units for which $\hbar=1$, one has $C_2= s(s+1)$ (or $s^2$ in the classical limit) but $s$ has now to be interpreted (for non-zero mass) as {\it superspin}.  Although this construction generalises to higher dimensions \cite{Pasqua:2004vq,Zumino:2004nb}, it appears that  its extension  to the other  super-Poincar\'e  Casimirs that become relevant in higher dimensions has not yet been explored (except for  a brief discussion specific to the 6D case \cite{Routh:2015ifa}).  In general, this extension will involve the intermediate construction of  supertranslation invariant  {\it even-rank} forms generalising the 2-form $P^pW_{mnp}(2)$. 

Whatever the merits of this approach, we think it desirable to have a construction of super-Poincar\'e Casimirs  that parallels the standard construction of Poincar\'e Casimirs.  Progress in this direction was first made, for the zero mass case,  by  Buchbinder and Kuzenko  \cite{Buchbinder:1998qv}. They suggested that the constraint $\Pslash Q=0$ should be imposed,  which is reasonable because it is implied by unitarity given $P^2=0$ \cite{Siegel:1988wk} and the two constraints are jointly supertranslation invariant.  If these constraints are assumed then $W(1)$ turns out to be supertranslation invariant; in fact, the  constraints imply that  
$W_m(1)= H P_m$ where $H$ is the ``classical superhelicity''\footnote{$H$ contains bi-linears of anticommuting variables; its eigenvalue in the quantum theory is the superhelicity shifted by $1/4$ \cite{Buchbinder:1998qv}; see also \cite{Arvanitakis:2016oyi}.}.  The same construction, again for zero mass, was later proposed, and generalized to higher spacetime dimensions,  by  Pasqua and Zumino \cite{Pasqua:2004vq,Zumino:2004nb}. 

Here we show how this Buchbinder-Kuzenko-Pasqua-Zumino  construction can be generalized to apply  to  massive superparticles.    Our method makes use of the fact that the universal enveloping algebra of the ${\cal N}=1$ super-Poincar\'e  algebra contains a BPS-saturated ${\cal N}=2$ super-Poincar\'e algebra, which is realized as a ``hidden'' symmetry algebra of massive superparticle actions \cite{Mezincescu:2014zba} (this is related to the ``off-shell symmetries''   of the {\it massless} superparticle \cite{Brink:1981rt,Bars:2003mp}).   For this larger ${\cal N}=2$ algebra, one can again impose a constraint on the supersymmetry charges that allows the construction of a  supertranslation invariant extension of the PL 3-form $W$  that is polynomial in super-Poincar\'e generators with  dimensionless coefficients;  we call it $Z$. Once again, the set of constraints required for the ${\cal N}=2$ super-translational invariance 
of $Z$ are implied by unitarity.
 
For zero mass the constraints on super-Poincar\'e generators reduce to $P^2=0$ and  $\Pslash Q=0$, and $Z$ reduces to the super PL 3-form $W(1)$.  
For non-zero mass we have a similar solution to the problem for an ${\cal N}=2$ BPS saturated 
super-Poincar\'e algebra, but the constraints on the two spinor charges allow one of them to be eliminated. This step yields
\begin{equation}\label{Yagain}
Z_{mnp} =  J_{[mn} P_{p]}+ \frac{i}{4m^2} \bar Q \Pslash \Gamma_{[mn} Q P_{p]} \, ,  
\end{equation}
which is, by construction, ${\cal N}=1$ super-translation invariant.  Given that  $P^2=-m^2$ for non-zero mass $m$, one may verify that
\begin{equation}
P^p Z_{mnp} = P^p W_{mnp}(2)\, . 
\end{equation}
From this fact, and the  expression (\ref{standard}) for the Casimir $C_2$, it follows that 
\begin{equation}
2m^4 C_2 =  9 P^p Z_{p[mn}P_{q]} Z^{mnq} =  3 P^2 Z_{mnp}Z^{mnp}\, , 
\end{equation}
where the last equality is a consequence of the identity
\begin{equation}\label{Zid}
Z_{[mnp} P_{q]} \equiv 0\, . 
\end{equation}
Using the mass-shell constraint again, we deduce that 
\begin{equation}
2m^2C_2 = -3 Z_{mnp}Z^{mnp} \, . 
\end{equation}
This shows that our construction of the Casimir of the  ${\cal N}=1$ super-Poincar\'e algebra yields the same result as the standard construction, but  
{\it in a way that  parallels the  non-supersymmetric case}.

Another purpose of this paper is to show how SPL tensors, in particular the 3-form $Z$, emerge naturally from a supertwistor formulation \cite{Ferber:1977qx} of massive superparticle mechanics. This is because, in the supertwistor formulation, (i) the constraints on the supertranslation charges required for supertranslation invariance of $Z$ become identities, and (ii)  for non-zero mass, {\it all}  supersymmetries of the action become manifest \cite{Mezincescu:2013nta}. 

The simplest  superparticle mechanics model  is due to Casalbuoni \cite{Casalbuoni:1976tz}  and  Brink and Schwarz \cite{Brink:1981nb}, and an action for
the 4D ${\cal N}=1$ Casalbuoni-Brink-Schwarz (CBS) superparticle of mass $m$  is 
\begin{equation}\label{masss}
S= \int\! dt \left\{ \Pi_t^m P_m - \frac{1}{2}e\left(P^2+m^2\right)\right\}\, , 
\end{equation}
where $e(t)$ is a Lagrange  multiplier for the mass-shell constraint, and  $\Pi^m_t$ is the pullback to the worldline, with arbitrary parameter $t$, 
of the supertranslation invariant superspace 1-form\footnote{The factor of $i$ here is due to the convention that complex conjugation inverts the order of anticommuting variables.}
\begin{equation}
\Pi^m = dX^m +i \bar\Theta\Gamma^m d\Theta\, . 
\end{equation}
The superspace coordinates comprise the Minkowski spacetime coordinates $X^m$ and  the {\it anticommuting} 4-component Majorana spinor $\Theta$, with Majorana 
conjugate $\bar\Theta$.  As the 4-momentum $P$ is also supertranslation invariant, the scalar Lagrangian is  super-Poincar\'e invariant.  For zero mass the CBS action is not strictly in Hamiltonian form because the  2-form  $\Omega = d(\Pi^m  P_m)$ is then non-invertible; this  is related to the existence  of a fermionic gauge invariance at zero mass \cite{Siegel:1983hh}.  
For non-zero mass,  the action (\ref{masss}) {\it is} in Hamiltonian form and $\Omega$ is the symplectic 2-form. The inverse of $\Omega$
determines the Poisson bracket (PB) of any two functions on the phase superspace. In particular, the non-zero PBs of the canonical variables are 
\begin{eqnarray}
\left\{X^m,P_n\right\}_{PB} &=& \delta^m_n\, , \qquad \left\{X^m,\Theta^\alpha\right\}_{PB} = -\frac{1}{2P^2} \left(\Pslash \Gamma^m\Theta\right)^\alpha\, ,  \\
\left\{\Theta^\alpha,\Theta^\beta\right\}_{PB} &=& \frac{i}{2P^2}  \left(\Pslash C\right)^{\alpha\beta}\, , \qquad 
 \left\{X^m,X^n\right\}_{PB} = -\frac{i}{2 P^2} \bar\Theta \Gamma^{mn}\Pslash \Theta\, . \nonumber
\end{eqnarray}
From the last of these relations we see that the quantum spacetime coordinates will not mutually commute, so the usual 
$P_m\to -i \partial_m$ rule for quantization is not applicable. As a result, covariant quantization is not straightforward even for non-zero mass. 

Supertwistor methods provide a way around this problem, as pointed out by Shirafuji for the massless  ${\cal N}=4$ CBS superparticle \cite{Shirafuji:1983zd}.  
They also allow a simple determination of the superspin content of  a quantum superparticle model.  This is because the introduction of (super)twistor variables introduces new gauge invariances that are associated with ``spin-shell'' constraints.  As the name suggests, these constraints determine  the (super)spin content because the constraint functions are simply related to the (super)PL
3-form.  It appears that a version of this relation was first noted in the context of particles in Anti-de Sitter space \cite{Cederwall:2004cf}. The  Minkowski space 
version has played a role previously in the context of particular  3D \cite{Mezincescu:2013nta} and 4D \cite{Fedoruk:2014vqa} massive particle  actions, and the relation of  6D  
super-PL tensors to the spin-shell constraints  of the massive 6D CBS superparticle was  one of the principal results of \cite{Routh:2015ifa}.   

What we wish to emphasize here is that the extension  from PL-tensors  to super-PL tensors becomes trivial in the (super)twistor formulation of (super)particle mechanics because  the relation of (super)PL tensors  to spin-shell constraints depends only on the  algebra of the constraints,  not on whether they are constraints for a particle or superparticle.   To illustrate this observation in a more generic setting, we  consider  a   novel 4D ``superspinning particle'' action inspired by the ``spinning particle''  \cite{Brink:1976sz,Brink:1976uf}; its supertwistor reformulation  shows that it describes, upon quantization,  the irreducible  4D ${\cal N}=1$ massive supermultiplet of superspin $1/2$. 

Finally, we unify the results relating (super-)PL tensors to  spin-shell constraints of 3D, 4D and 6D (super)particle mechanics  by means of an $Sl(2;\bK)$ bi-spinor notation \cite{Arvanitakis:2016vnp}, where
$\bK=\bR,\bC,\bH$ (the associative normed division algebras). This makes use of the  relation  of supersymmetric field theories in Minkowski spacetimes of dimension $d=2+ {\rm dim}\, \bK$ to the normed division algebras $\bK=\bR,\bC,\bH,\bO$   \cite{Kugo:1982bn,Sudbery,Evans:1987tm,Baez:2009xt}, although we have not yet seen how to use the $\bK=\bO$ case of this relation to extend our 
(S)PL  tensor results to 10D.

\section{4D Super-Pauli-Lubanski}\label{sec:4D}

 For simplicity, we shall assume that the Poincar\'e charges are realized as 
functions  on  phase space, so that the Lie product  is the Poisson bracket and the associative product of  the enveloping algebra
is just the product of functions. The non-zero PB relations of the Poincar\'e charges are
\begin{equation}\label{Poinc}
\left\{J_{mn}, J_{pq}\right\}_{PB} = 2\eta_{p[m} J_{n]q} - 2\eta_{q[m}J_{n]p} \, , \qquad 
\left\{J_{mn},P_p\right\}_{PB} = 2\eta_{p[m} P_{n]}\, .
\end{equation}
 
Our first goal is to find a supertranslation invariant SPL  3-form in the context of an ${\cal N}=1$  super-Poincar\'e algebra
spanned by the Lorentz generators $J_{mn}$ and the supertranslation  generators $(P_m,Q^\alpha)$, where
$Q^\alpha$ are the components of a minimal spinor. We assume, for simplicity of presentation,  that the minimal spinor is Majorana, as it is in 4D (in which case
 $\alpha=1,2,3,4$)  but otherwise there is no restriction on the spacetime dimension.  We also continue to assume that the  Lie product is  a Poisson bracket,  now 
 suitably generalized to accomodate anticommuting functions; in this case the components of $Q$  are mutually anticommuting and their Poisson brackets are symmetric, rather than antisymmetric, under interchange. The additional non-zero PB relations defining the  ${\cal N}=1$ super-Poincar\'e algebra are 
\begin{equation}\label{SPPBs}
\left\{J_{mn},Q_\alpha\right\}_{PB} =  \frac{1}{2} \left(\Gamma_{mn} Q\right)_\alpha \, ,  \qquad 
\left\{ Q_\alpha,Q_\beta\right\}_{PB}  = -i \left(\Pslash C\right)_{\alpha\beta}\, .  
\end{equation}
We recall that $C$ is the charge conjugation matrix. Given the restriction we have imposed on the spacetime dimension, the matrix $C$ is antisymmetric and 
the matrices $\Gamma_m C$ are symmetric. 

As remarked in the introduction, there is no ${\cal N}=1$ supertranslation invariant extension of the PL 3-form (\ref{PL3form})  with purely numerical (dimensionless) coefficients
unless one imposes the (supertranslation invariant) conditions $P^2=0$ and $\Pslash Q=0$, but then we are restricted to massless representations. To generalize this idea
to  massive representations, for which $P^2=-m^2$ for $m\ne0$, we introduce the new supersymmetry charge $\tilde Q$ by the relation $\Pslash Q = m\tilde Q$. 
A computation of the PB relations obeyed by $\tilde Q$ yields
\begin{equation}
\left\{ \tilde Q_\alpha,\tilde Q_\beta\right\}_{PB}  = -i \left(\Pslash C\right)_{\alpha\beta}\, , \qquad  \left\{ Q_\alpha, \tilde Q_\beta\right\}_{PB} = -im C_{\alpha\beta}\, . 
\end{equation}
These relations confirm  that $\tilde Q$ is a second supercharge, and they also show that the mass $m$ is a central charge of the resulting ${\cal N}=2$ supersymmetry algebra.  As we explain in a subsection to follow, it is actually the largest central charge compatible with the BPS unitarity bound of the quantum theory. 

To summarize: we  have a generalization of the zero-mass  BK constraints to non-zero mass $m$, but now in the context of the BPS ${\cal N}=2$ algebra.  These constraints are
\begin{equation}\label{constraints}
P^2+m^2=0\, , \qquad \Pslash Q - m\tilde Q =0 \quad \left(\Rightarrow\,  \Pslash \tilde Q + m Q =0\right). 
\end{equation}
We now seek an ${\cal N}=2$ supertranslation invariant extension of the PL 3-form (\ref{PL3form}).  It is not difficult to show that the 3-form 
\begin{equation}\label{Z}
Z_{mnp} = J_{[mn}P_{p]} - \frac{i}{24} \left(\bar Q\Gamma_{mnp} Q + \bar{\tilde Q} \Gamma_{mnp} \tilde Q\right) 
\end{equation}
has this property.  Its Poisson bracket  with $P$ is obviously zero, and 
\begin{eqnarray}
\left\{Z_{mnp}, Q_\alpha\right\}_{PB} &=& \frac{1}{12} \left[\Gamma_{mnp}\left(\Pslash Q- m\tilde Q\right)\right]_\alpha =0\, , 
\nonumber \\
\left\{Z_{mnp}, \tilde Q_\alpha\right\}_{PB} &=& \frac{1}{12} \left[\Gamma_{mnp}\left(\Pslash \tilde Q + m Q\right)\right]_\alpha=0\, . 
\end{eqnarray}
If we use the relation $m\tilde Q = \Pslash Q$ to eliminate $\tilde Q$ from  the expression (\ref{Z}),  we find that
\begin{equation}\label{Yagain2}
Z_{mnp} =  J_{[mn} P_{p]}+ \frac{i}{4m^2} \bar Q \Pslash \Gamma_{[mn} Q P_{p]} \, ,  
\end{equation}
which is the result stated in the Introduction.  This is still ${\cal N}=2$ supertranslation invariant (because the constraints are ${\cal N}=2$ supertranslation invariant) and hence
${\cal N}=1$ supertranslation invariant. 

We have shown in the Introduction how the SPL tensor $Z$ is related to the tensor $W(2)$ used in the standard construction of the super-Poincar\'e spin Casimir. 
We used there the fact that $Z_{[mnp} P_{q]} \equiv 0$, which implies that
\begin{equation}\label{Fink}
Z_{mnp} = {\cal U}_{[mn} P_{p]}
\end{equation}
for some 2-form ${\cal U}$. Clearly, we may add to  ${\cal U}$ the exterior product of $P$ with any 1-form, but this ambiguity is eliminated if we require that 
\begin{equation}\label{Fink2}
P^m {\cal U}_{mn}\equiv 0\, . 
\end{equation}
In this case, 
\begin{equation}\label{Fink3}
{\cal U}_{mn} = J_{mn}  -\frac{2}{m^2} P_{[m}J_{n]q}P^q + \frac{i}{4m^2}\bar Q\Pslash \Gamma_{mn} Q  \, .
\end{equation}
This expression was originally found by Finkelstein and Villasante \cite{Finkelstein:1984iu}. The constraint (\ref{Fink2}) 
implies that only the space components of ${\cal U}$ are non-zero in the rest frame.  Notice too that
\be
P^pZ_{mnp}=-\frac{m^2}{3} {\cal U}_{mn}\, , 
\ee
which confirms that the Casimir $C_2$ is proportional to $|{\cal U}|^2$. This construction generalizes to all spin Casimirs of 
super-Poincar\'e groups in any higher spacetime dimension  \cite{Kwon:1987xc}. 

\subsection{Quantum unitarity constraints}

The quantum (anti)commutation relations for the operator charges spanning the super-Poincar\'e algebra can be obtained from the PB relations used above
by the usual procedure of replacing the PB by $-i$ times the (anti)commutator. For the ${\cal N}=1$ super-Poincar\'e algebra with charge $Q$ we then have 
the anticommutation relations
\begin{equation}
\{Q_\alpha, Q_\beta\} = (\Pslash C)_{\alpha \beta} \, . 
\end{equation}
Recall that we have restricted our discussion, for simplicity of presentation, to those spacetime dimensions for which  $Q$ is a Majorana spinor. We may then choose a real basis for the Dirac matrices in  which $C=\Gamma^0$.  Majorana spinors are real in such a  basis, so it would be natural to suppose  that the quantum operator $Q$ should be Hermitian. 
However, the classical $Q$ cannot actually be ``real''  because it is anticommuting, and because of this one should rather suppose that the quantum operator 
$Q$ is {\it either}  Hermitian {\it or}
anti-Hermitian\footnote{The product $H=2i\mu\nu$ is ``real'' for anticommuting ``real'' $\mu$ and $\nu$, and this becomes $\hat H= i[\hat\mu,\hat\nu]$ for the corresponding 
quantum operators, but hermiticity of $\hat H$ allows the operators $(\hat\mu,\hat\nu)$ to be either both Hermitian or both anti-Hermitian.}. As we shall now see, supersymmetry correlates
this choice with the sign of the energy, which is fortunate since both positive and negative energies are needed for second quantization. 

Given that $C=\Gamma^0$, and choosing  the rest-frame for a massive particle, we have
\begin{equation}
\left\{Q_\alpha,Q_\beta\right\} = P^0 \delta_{\alpha\beta} \quad \Rightarrow \quad 2Q_\alpha^2 = P^0 \qquad ({\rm no\ sum}). 
\end{equation}
Taking the expectation value in any state $|\Psi\rangle$ we deduce that 
\begin{equation}
2\| Q_\alpha|\Psi\rangle \|^2 = \pm |P^0|  \qquad ({\rm no\ sum}), 
\end{equation}
where the top sign is for hermitian $Q$ and the bottom sign for anti-hermitian $Q$. Assuming the absence of negative norm states, i.e. assuming unitarity, we deduce that $Q$ is hermitian for positive energy and anti-hermitian for negative energy.  However, for what follows we assume that $P^0>0$ and that $Q$ is hermitian. 

Now we turn to the  ${\cal N}=2$ super-Poincar\'e algebra with supercharges $(Q,\tilde Q)$. Relabeling these supercharges
as $Q^i$ ($i=1,2$), we have  the anticommutation relations
\begin{equation}
\{Q^{i}_\alpha, Q^{ j}_\beta\} = \delta^{ij} (\Pslash C)_{\alpha \beta} + z \epsilon^{ij} C_{\alpha \beta}\, . 
\end{equation}
Here we allow for arbitrary real central charge $z$, although $z=m$ for the ${\cal N}=2$ algebra deduced from ${\cal N}=1$ superparticle mechanics. 
Using these relations, one may show that 
\begin{equation}
\left\{\left(\Pslash Q -m\tilde Q\right)_\alpha , \left(\Pslash Q -m\tilde Q\right)_\beta\right\} = 2m(m-z) \left(\Pslash C\right)_{\alpha\beta}\, . 
\end{equation}
Again choosing  a Dirac matrix basis such that $C=\Gamma^0$, and the rest-frame for a massive particle, we deduce that 
\begin{equation}
\|\left(\Pslash Q - m\tilde Q\right)_\alpha|\Psi\rangle \|^2 = m(m-z) P^0\, . 
\end{equation}
Since $P^0=m>0$,  we see that unitarity requires $z\le m$.  When this ``BPS bound'' is saturated, i.e. when $z=m$, the operator 
$\Pslash Q - m\tilde Q$ has zero norm in any state. Assuming the absence of zero-norm states, we deduce that 
\begin{equation}
\left(\Pslash Q - m\tilde Q\right)_\alpha |\Psi\rangle =0\, , 
\end{equation}
for any state $|\Psi\rangle$.  Classically,  this becomes the additional constraint $\Pslash Q = m\tilde Q$ of (\ref{constraints}) that we used in the construction of the super-PL pseudo-vector for particles of mass $m$. 

\section{Massive superparticles and supertwistors}\label{sec:sparticle}\label{sec:three}

We now aim to show how the above construction of a super-PL 3-form emerges naturally from a supertwistor formulation of massive 
superparticle mechanics.  To do so it is simplest to first replace 4-component Majorana spinors by two-component Weyl spinors.
Specifically,  the anticommuting Majorana spinor $\Theta$ becomes the complex $Sl(2;\bC)$ doublet $\Theta^A$ with complex conjugate 
$\Theta^{A'}$ ($A,A'=1,2$) and the position  4-vector becomes the hermitian bi-spinor $X^{AA'}$, with canonically conjugate 4-momentum
$P_{AA'}$.  The 4D  CBS superparticle action (\ref{masss}) in this notation is\footnote{See  \cite{Mezincescu:2015apa} for details of the conversion from Lorentz-vector
notation in our conventions.}  
\begin{eqnarray}\label{CBS-NP}
S = \int\! dt \left\{-\frac{1}{2} \Pi_t ^{AA'} P_{AA'}  - \frac{1}{2} e\left(P^2 +m^2 \right) \right\}\, ,
\end{eqnarray}
where
\begin{equation}
\Pi_t^{AA'} = \dot X^{AA'} + i\left( \Theta^{A'} \dot\Theta^{A} - \dot\Theta^{A'}\Theta^A\right) 
\end{equation}
and 
\begin{equation}\label{vecnorm}
P^2 = -\frac{1}{2}P^{AA'}P_{AA'} \, , \qquad P^{AA'} = \varepsilon^{AB} \varepsilon^{A'B'} P_{BB'}\, . 
\end{equation}

Next, we express $P_{AA'}$ in terms of an $SU(2)$ doublet  of (commuting) Weyl spinors 
$U_A{}^I$ ($I=1,2$),  with complex conjugates $U_{A'\, I}$,  as follows: 
\begin{equation}
P_{AA'} = \mp U_A{}^I U_{A'\, I} \, . 
\end{equation}
The top (bottom) sign corresponds to the choice of positive (negative) energy.  The mass-shell constraint is now 
\begin{equation}\label{newms}
0= \varphi := |\det U|^2 -m^2 \, , 
\end{equation}
where $U$ is the complex $2\times 2$ matrix with entries $U_A{}^I$.  

Substitution also yields 
\begin{equation}
-\frac{1}{2} \Pi_t ^{AA'} P_{AA'} =  \dot U_A{}^I W_I{}^A + \dot U_{A'\, I} W^{I\, A'} \pm i \bar\mu_I \dot\mu^I + \frac{d}{dt}(\cdots)\,  , 
\end{equation}
where 
\begin{equation}
\mu^I = \Theta^A U_A{}^I \, , \qquad \bar\mu_I = \Theta^{A'} U_{A'\, I}\, , 
\end{equation}
and 
\begin{equation}
W_I{}^A = \mp\frac{1}{2} \left[ X^{AA'} U_{A'\, I} + i \bar\mu_I\Theta^A\right] \, . 
\end{equation}
This last expression (together with its complex conjugate) implies the identity
\begin{equation}\label{CBS-id}
0 \equiv G_0^I{}_J := U_A{}^I W_J{}^A - U_{A'\, J} W^{I\, A'} \mp i \mu^I\bar\mu_J\, . 
\end{equation}
Notice that $G_0^I{}_J$ are the entries of an {\it anti-hermitian} matrix. Its trace is 
\begin{equation}
G_0 := G_0^I{}_I = U_A{}^I W_I{}^A - U_{A'\, I} W^{I\, A'} \mp i\mu^I\bar\mu_I\, . 
\end{equation}

The identity (\ref{CBS-id}) ceases to be an identity if  $W_I{}^A$ is interpreted as an {\it independent} variable canonically conjugate to $U_A{}^I$, so this interpretation requires us to 
 impose the equations $G_0^I{}_J=0$ as constraints by means of Lagrange multipliers. Taking into account the mass-shell constraint  $\varphi=0$, we thus arrive at the 
equivalent action
\begin{equation}\label{equivtwist}
S= \int \! dt\left\{ \dot U_A{}^I W_I{}^A + \dot U_{A'\, I} W^{I\, A'} \pm i \bar\mu_I \dot\mu^I - s^J{}_I G_0^I{}_J - \rho \varphi \right\}\, , 
\end{equation}
where $s^J{}_I$ and $\rho$ are Lagrange multipliers for $4+1=5$ first-class constraints. The gauge invariance generated by $\varphi$ is equivalent to a time reparametrization\footnote{It differs by a ``trivial'' gauge transformation; see \cite{Routh:2015ifa} for a discussion of this point.}.

The above is  a summary of  the appendix to \cite{Mezincescu:2015apa}  expressed in a slightly different notation. Some further details 
may be found there; in particular the Poisson bracket relations, which may be used to show that the constraints $G_0^I{}_J$ span the Lie algebra $U(2)$ with respect to 
Poisson brackets. As the mass-shell constraint is manifestly $U(2)$ invariant, all five constraints are first-class and hence generate gauge invariances. The variables 
$(U_A{}^I,W_I{}^A; \mu^I)$ may be viewed, for each $I=1,2$, as a $(4|1)$-plet of $SU(2,2|1)$, which is a cover of the
 ${\cal N}=1$ 4D superconformal group. In other words, the phase space is parametrized by a pair of supertwistors, and only the mass-shell term breaks the 
 $SU(2,2|1)$ invariance.  
 
In addition to its worldline diffeomorphism and $U(2)$ gauge invariances,  the action (\ref{equivtwist}) is ${\cal N}=1$ super-Poincar\'e invariant. The Lorentz charges are 
\begin{eqnarray}
J_A{}^B &=& U_A{}^I W_I{}^B- \frac{1}{2}\delta_A^B\left(U_C{}^KW_K{}^C\right)\, , 
\end{eqnarray}
and complex conjugates.  The anticommuting variables do not appear here because they are now Lorentz scalars; this is one of the simplifying features of 
the supertwistor formulation.  The supersymmetry spinor charge (and complex conjugate) is 
\begin{equation}
Q_A = \mp U_A{}^I \bar \mu_I \, , \qquad Q_{A'} = \mp U_{A'\, I} \mu^I\, . 
\end{equation}
However, there is a further ``hidden'' supersymmetry \cite{Mezincescu:2014zba}, with spinor charge (and complex conjugate) 
\begin{equation}
\tilde Q_A = \frac{\det \bar U}{m} \, U_A{}^I\mu_I \, , \qquad {\tilde Q}_{A'} =   -\frac{\det U}{m}  U_{A'\, I} \bar \mu^I\, . 
\end{equation}
Notice that these charges satisfy the identity
\begin{equation}\label{Qsid}
P_{AA'} Q^{A'} + m \tilde Q_A\equiv 0\, . 
\end{equation}
The non-zero Poisson brackets of both  supercharges are
\begin{eqnarray}
\left\{Q_A, Q_{A'}\right\}_{PB} &=& iP_{AA'} \, , \qquad \left\{\tilde Q_A, \tilde Q_{A'}\right\}_{PB}= iP_{AA'}\, ,  \nonumber \\
\left\{Q_A, \tilde Q_B\right\}_{PB} &=& im\,  \varepsilon_{AB}\, , 
\quad \left\{Q_{A'}, \tilde Q_{B'}\right\}_{PB} =  im\,  \varepsilon_{A'B'}\, . 
\end{eqnarray}
We see that the ${\cal N}=2$ supersymmetry algebra has a central charge, and it follows from the identity (\ref{Qsid}) that it saturates the BPS bound. 

We now turn to the super-PL 3-form $Z$ as given in (\ref{Fink}),  with ${\cal U}$ as given in (\ref{Fink3}).  In Weyl  spinor notation (\ref{Fink})  becomes 
\begin{equation}\label{ZAAP}
Z_{AA'} =  \left({\cal U}_A{}^B P_{BA'} - {\cal U}_{A'}{}^{B'} P_{AB'}\right)\, .  
\end{equation}
Both terms on the right hand side contribute equally as a consequence of  (\ref{Fink2}) so we may simplify
this formula to 
\begin{equation}\label{ZWeyl}
Z_{AA'} = 2 {\cal U}_A{}^B P_{BA'} \, ,  
\end{equation}
where
\begin{equation}
{\cal U}_A{}^B = J_A{}^B - \frac{1}{m^2} P^{BC'} J_{C'}{}^{D'}P_{AD'} -
\frac{i}{2m^2} \left(Q_A P^{BC'}Q_{C'}+Q^BP_{A}{}^{C'} Q_{C'}\right)\, , 
\end{equation}
which is (\ref{Fink3}) in Weyl spinor notation. Notice that $Z$ is now represented by an {\it anti-hermitian} matrix\footnote{Multiplication by $i$ would yield an Hermitian matrix  but this would be less natural, for reasons to be explained in the following section.}.  When this matrix is expressed in terms of supertwistor variables, one finds that 
\begin{equation}\label{SPL-CBS}
Z_{AA'} =  \mp  U_A{}^J U_{A'\, I} \left(G_0^I{}_J - \frac{1}{2} \delta^I_J G_0\right)\, . 
\end{equation}
That is,  $Z$ is the Lorentz tensor associated to the triplet of $SU(2)$  spin-shell constraint functions of the massive 4D CBS superparticle action, and these constraints tell us that 
the super-PL  3-form is zero and hence that the quantum superparticle associated to this action has zero superspin (this motivates the zero subscript on the spin-shell constraint functions).  

A curiosity of this 4D case is that there is also a $U(1)$ ``spin-shell'' constraint that has no direct relation to spin.  One might be concerned about the possibility of a global $U(1)$ anomaly due to the ``worldline fermions''  \cite{Elitzur:1985xj,Arvanitakis:2016oyi}  but there is no anomaly here because the number of fermi oscillators is even.

\subsection{The superspinning particle}

Now we generalize by adding, to the action (\ref{CBS-NP}),  terms that are bilinear in additional anticommuting variables:  a Lorentz vector $\lambda^{AA'}$ and a scalar $\xi$.
This new ``superspinning particle'' action  is 
\begin{eqnarray}\label{Spartact-twist}
S &=& \int\! dt \Big\{-\frac{1}{2} \Pi_t ^{AA'} P_{AA'}    - 
\frac{i}{4} \lambda^{AA'} \dot \lambda_{AA'} + \frac{i}{2} \xi\dot\xi \nonumber \\
&& \ - \frac{1}{2} e\left(P^2 +m^2 \right) + \frac{i}{4} \zeta\left(\lambda^{AA'}P_{AA'} -2m\, \xi\right) \Big\}\, , 
\end{eqnarray}
where $\zeta$ is a new anticommuting Lagrange multiplier for a new constraint; the new constraint function generates a local worldline supersymmetry (exactly as it does for the massive spinning particle of 
\cite{Brink:1976uf} because the $\Theta$-dependent terms are invariant under this new gauge transformation).  

If the mass is set to zero and the anticommuting scalar variable $\xi$ is omitted then  we get the ``spinning superparticle'' of \cite{KowalskiGlikman:1987pw,Bergshoeff:1989uj}.  We are  thus considering a very simple extension to non-zero mass of the spinning superparticle. A much more complicated ``massive spinning superparticle'' action 
was proposed in \cite{KowalskiGlikman:1989eb} but we postpone comment on this to our concluding discussion.  

Now we set
\begin{equation}
P_{AA'}= \mp U_A{}^I U_{A'\, I} \, , \qquad \lambda^{AA'} = \frac{1}{m} \left[U_A{}^I U_{A'\, J} \,  \bfs^J{}_I \cdot \bfpsi + P_{AA'}\,  \xi\right] \, , 
\end{equation}
where $\bfs$ is the triplet of hermitian Pauli-matrices and $\bfpsi$ a triplet of ``real'' anticommuting variables. The constraints are solved by this substitution provided that
we impose the new mass shell constraint (\ref{newms}), and substitution yields 
\begin{equation}
-\frac{1}{2} \Pi_t^{AA'} P_{AA'} = \dot U_A{}^I W_I{}^A + \dot U_{A'\, I}  W^{I\, A'} \pm i\bar\mu_I\dot\mu^I
+ \frac{i}{2}\bfpsi\cdot \dot\bfpsi + \frac{d}{dt} \left(\cdots\right)\, , 
\end{equation}
but now with  
\begin{equation}\label{defsW}
W_I{}^A = \mp\frac{1}{2} \left(X^{AA'}\,  U_{A'\, I} +i \bar\mu_I \theta^A\right) 
 \mp \frac{i\det \bar U}{2m^2}  \varepsilon^{AB}U_B{}^J  \varepsilon_{JK}\, \bfs^K{}_I\cdot    \left(\bfpsi \xi \pm \bfS \right) \, , 
\end{equation}
where $\bar U$ is the complex conjugate of the matrix $U$ and 
\begin{equation}
\bfS = -\frac{i}{2} \bfpsi\times \bfpsi \, . 
\end{equation}
The identity (\ref{CBS-id}) is now modified to 
\begin{equation}\label{modss}
0\equiv G^I{}_J = G_0^I{}_J - i\bfs^I{}_J \cdot \bfS\, , 
\end{equation}
where $G_0^I{}_J$ is the matrix of spin-shell constraint functions of (\ref{CBS-id}).  This is the same as the spin-shell constraint found in  \cite{Mezincescu:2015apa} for the
massive spinning particle except that $G_0^I{}_J$ now includes a term quadratic in the anticommuting variables $\mu^I$.  Notice that the other anticommuting 
variables  appear only in the traceless part of  the matrix $G^I{}_J$, so its trace ($G$) equals $G_0$. 

As for the CBS superparticle, we may interpret $W_I{}^A$ as the set of complex variables canonically conjugate to $U_A{}^I$ by imposing the equations $G^I{}_J=0$ 
as constraints via Lagrange multipliers. We thus find the following equivalent version of the superspinning particle action:
\begin{equation}\label{newform}
S= \int dt\left\{ \dot U_A{}^I W_I{}^A+ \dot U_{A'\, I} W^{I\, A'} \pm i\bar\mu_I\dot\mu^I + \frac{i}{2}\bfpsi\cdot \dot{\bfpsi} - 
s^J{}_IG^I{}_J - \rho \varphi\right\}\, . 
\end{equation}
The non-zero Poisson brackets of the canonical variables are
\begin{eqnarray}
\left\{U_A{}^I, W_J{}^B\right\}_{PB} &=& \delta^B_A \delta^I_J\, , \qquad  
\left\{U_{A'\, I} , \bar W^{J\, B'}\right\}_{PB} = \delta^{B'}_{A'} \delta_I^J\, , \nonumber\\
\left\{\mu^I ,\bar \mu_J\right\}_{PB} &=& \mp \, i \delta^I_J\, , \qquad \left\{\psi_i,\psi_j\right\}_{PB} =-i \delta_{ij}\, ,  
\end{eqnarray}
where $\psi_i$ ($i=1,2,3$) are the components of $\bfpsi$.  All constraints are first class and they generate gauge transformations of the action. 
The constraint functions  $G^I{}_J$ generate a local $U(2)$ invariance, just as they did for the CBS superparticle. 

Using (\ref{modss}) we may rewrite the relation  (\ref{SPL-CBS}) between the SPL tensor $Z$ and the CBS massive superparticle constraint functions as
\begin{equation}
Z_{AA'} \pm  iU_A{}^J U_{A'\, I} \, \bfs^I{}_J  \cdot\bfS = \mp  U_A{}^J U_{A'\, I} \left(G^I{}_J - \frac{1}{2} \delta^I_J G\right)\, . 
\end{equation}
The additional $\bfS$-dependent term on the left hand side cancels with the same term on the right hand side coming from the $\bfS$-dependence of the  traceless part of  the spin-shell constraint matrix $G^I{}_J$. These $SU(2)$ spin-shell constraints now tell us that 
\begin{equation}\label{result}
Z_{AA'} = \mp i U_A{}^J U_{A'\, I} \, \bfs^I{}_J \cdot\bfS \, . 
\end{equation}
This is essentially the same result as that found in  \cite{Mezincescu:2015apa} for the massive spinning particle,  
but now it is a result for the SPL  3-form $Z$ rather than the  PL  3-form $W$.  

Passing to the quantum theory we have\footnote{There is a sign difference in the norm of $Z$ relative to that of $P$ in (\ref{vecnorm}) because $Z$ is anti-hermitian rather than hermitian.}
\begin{equation}
|\hat Z|^2 =  \frac{1}{2} Z^{AA'} Z_{AA'} = m^2 |\hat{\bfS}|^2\, , 
\end{equation}
where the second equality uses the mass-shell constraint.  As  explained in detail in \cite{Mezincescu:2015apa}, where it was used  to confirm that the spinning particle has spin $\tfrac{1}{2}$, 
the operator $\hat\bfS$ is such that 
\begin{equation}
 |\hat{\bfS}|^2 = \tfrac{3}{4}\, . 
 \end{equation}
In the present context this implies that the superspinning particle has  superspin $\tfrac{1}{2}$. That is, it describes a particle supermultiplet with eight
polarization states:  the three helicity states of a spin-$1$ particle, the four helicity states of two spin-$\frac{1}{2}$ particles,  and two spin-$0$ states.

\section{(S)PL tensors for dimensions $d=3,4,6$}

We have seen that the spin-shell constraints arising in the (super)twistor formulation of 4D massive (super)particle dynamics are directly related to the (super)PL 3-form
that determines the (super)particle's (super)spin. These results complement those found for the 6D CBS superparticle in  \cite{Routh:2015ifa}.  In fact, the relation of (super-)PL tensors to spin-shell constraints arising  in the (super)twistor formulation of the 4D and 6D (super)particle can be understood in a unified way that  deals simultaneously with the 3D, 4D and 6D cases. This is made possible by the observation that the Lorentz group for Minkowski spacetime of dimension $d=3,4,6$ is 
$Sl(2;\bK)$ for $\bK=\bR,\bC,\bH$, the three associative normed division algebras over the reals \cite{Kugo:1982bn}, and the related observation that the conformal group in these dimensions is  
$Sp(4;\bK)$ \cite{Sudbery}.  Here we shall follow a recent application of these 
observations to the (super)twistor formulation of (super)particle mechanics \cite{Arvanitakis:2016vnp},  initially focusing on the non-supersymmetric case. 

Minkowski coordinates in dimension $d=2+ {\rm dim}\, \bK$  correspond to entries of a $2\times2$ hermitian matrix $\bX$ over $\bK$, and the transformation 
 \begin{equation}
\bX \to \bL \bX \bL^\dagger\,, \qquad \det (\bL\bL^\dagger)=1
\end{equation}
is a Lorentz transformation, although it includes an additional $U(1)$ transformation in the $\bK=\bC$ case because the unit determinant condition is on $\bL\bL^\dagger$ rather than $\bL$; this 
 is natural in the current  context since the determinant of a quaternionic matrix is intrinsically defined only if it is (quaternionic) Hermitian.  
  
If $\bX(t)$ represents the position of a particle at parameter time $t$ on its worldline, then an hermitian matrix $\bP(t)$ represents the particle's $d$-momentum but with Lorentz transformation 
\begin{equation}
\bP \to (\bL^\dagger)^{-1} \bP \bL^{-1}\, . 
\end{equation}
We may therefore get a Poincar\'e invariant from the matrix product $\dot \bX \bP$ by taking the real part of its trace, which we shall call the ``real-trace'' and denote by $\tr_\mathbb{R}$; 
the real-trace has the cyclicity property $\tr_\mathbb{R}(\bA\bB\bC) = \tr_\mathbb{R}(\bC\bA\bB)$ even for quaternionic matrices. 

We may get another Poincar\'e invariant by taking the determinant of $\bP$. We choose a normalisation of $\bP$, and a ``mostly plus'' Minkowski metric convention, such that 
\begin{equation}
\det \bP = -p^2 \, . 
\end{equation}
For a similar normalisation of $\bX$, the standard manifestly Poincar\'e-invariant phase-space action for the relativistic point particle of mass $m$ becomes
\begin{equation}
S= \int\! dt \left\{\frac{1}{2}\tr_\mathbb{R} \left(\dot\bX \bP\right) - \frac{1}{2}e \left(\det\bP -m^2\right)\right\}\, , 
 \end{equation}
 where $e(t)$ is a Lagrange multiplier for the mass-shell constraint. 
 
Now we write
\begin{equation}\label{LtoR}
\bP = \mp \bU\bU^\dagger \, , 
\end{equation}
where the top/bottom sign is for positive/negative energy,  and $\bU$ is a $2\times 2$ matrix subject to  the transformations
 \begin{equation}
\bU(t) \to ({\bL}^\dagger)^{-1} \bU(t) \bN(t) \, , \qquad \bN\bN^\dagger = \bI\, , 
\end{equation}
where ${\bN}(t)$ parametrises a map from the particle's worldline to the rotation group $O(2;\bK)$; this is defined to preserve a $\bK$-hermitian quadratic form on $\bK^2$, so that
\begin{equation}
O(2;\bR) \cong  O(2)\, , \qquad O(2;\bC) \cong U(2)\, , \qquad O(2;\bH)\cong {\rm Spin}(5)\, . 
\end{equation}
Notice that the ``rotation'' group for $\bK=\bC$ has an additional $U(1)$ factor, consistent with the additional $U(1)$ factor in the $d=4$ ``Lorentz'' group. 

Substitution for $\bP$ now yields the new mass-shell constraint
\begin{equation}\label{twistmass}
\det (\bU\bU^\dagger) = m^2\, . 
\end{equation}
In addition, 
\begin{equation}
\frac{1}{2}\tr_\mathbb{R} (\dot\bX \bP)  = \tr_\mathbb{R} (\dot\bU \bW^\dagger) + \frac{d}{dt}\left(\cdots\right)\, , 
\end{equation}
where
\begin{equation}\label{incidence}
\bW = \pm \bX \bU\, . 
\end{equation}
This ``incidence relation''  implies the identity 
\begin{equation}
0 \equiv \bG_0 := \bU^\dagger\bW - \bW^\dagger \bU\, . 
\end{equation}
In order to interpret the $2\times2$ matrix $\bW$ as canonically conjugate to $\bU$ we drop the incidence relation and impose $\bG_0=0$ as a constraint  with an anti-hermitian Lagrange multiplier $\bS$. This yields the action 
\begin{equation}\label{newaction}
S = \int\! dt\left\{ \tr_\mathbb{R} \left(\dot\bU \bW^\dagger\right) - \tr_\mathbb{R} \left(\bS \bG_0\right) - \ell \left[\det(\bU\bU^\dagger)-m^2\right]\right\}\, .
\end{equation}
This action is Poincar\'e invariant with Noether charges 
\begin{equation}
\bP= \mp \bU\bU^\dagger\, , \qquad \bJ = \bU\bW^\dagger - \frac{1}{2} \tr_\mathbb{R}(\bU\bW^\dagger) \bI\, . 
\end{equation}
The anti-hermitian matrix constraint function $\bG_0$ is the generator of an $O(2;\bK)$ gauge transformation. In particular, $\bG_0$ itself transforms by conjugation 
with an element $\bN$ of $O(2;\bK)$
\begin{equation}
\bG_0 \to \bN^\dagger\bG_0 \bN\, \qquad (\bN^\dagger \bN =\bI).
\end{equation}

With the exception of the mass-shell constraint term, the rest of the action is invariant under the larger  group $Sp(4;\bK)$, defined to preserve a skew-hermitian 
quadratic form on $\bK^4$. This is the conformal group of $d$-dimensional Minkowski spacetime for $d=2+{\rm dim}\, \bK$, except that $Sp(4;\bC)\cong U(2,2)$, 
which implies that there is again an additional $U(1)$ factor for $d=4$.  The $4$-plet of $Sp(4;\bK)$ is equivalent to a twistor (a spinor of the conformal group) 
and the $4\times 2$ matrix
\begin{equation}
{\cal Z}  = \left(\begin{array}{c} \bU \\ \bW\end{array}\right)
\end{equation}
constitutes a pair of twistors, acted upon from the left by $Sp(4;\bK)$ and from the right by the $O(2;\bK)$ gauge group.  

The above is a summary of some results of \cite{Arvanitakis:2016vnp}, which we now use to investigate  (S)PL tensors.

\subsection{Spin-shell constraints and the quadratic Casimir}

Because the Poincar\'e Noether charges of the action (\ref{newaction})  are gauge invariant,  they have zero Poisson brackets with the spin-shell constraint functions $\bG_0$. 
It follows that these constraint functions are translation invariant.  As the matrix $\bU$ is also translation invariant, it also follows that 
the Lorentz tensor  
\begin{equation}
\bZ_+= \pm \bU \bG_0 \bU^\dagger\, 
\end{equation}
is translation invariant and hence represents a PL  tensor if it can be re-expressed in terms of the Poincar\'e Noether charges. 
Substitution for $\bG$ yields
\begin{equation}
\bZ_+   = \bJ\bP - \bP\bJ^\dagger \, , 
\end{equation}
showing that $\bZ_+$ is indeed a PL tensor.  Notice that $\bZ_+$ is anti-hermitian, which implies that it has $(3\, {\rm dim}\bK -2)$ independent real components. It is equivalent to a Lorentz pseudo-scalar for  $d=3$ and  a Lorentz pseudo-vector for $d=4$. For $d=6$ it is equivalent to a Lorentz 3-form that is either self-dual or anti-self-dual, and we may suppose it to be self-dual. 

In general, if  a Lorentz vector $h$ is represented by an Hermitian matrix $\bH$ transforming as $\bP$ then 
\begin{equation}
\tilde \bH = \bH - \tr_{\mathbb{R}}(\bH)
\end{equation}
is the hermitian matrix representing the corresponding co-vector  \cite{Schray:1994fc}; i.e. it transforms as $\bX$. This follows from the identity
$\bH\tilde\bH = h^2 \bI$ \cite{Baez:2010ye}.  Applying this result to $\bP$ we have.
\begin{equation}
\tilde \bP = \bP- (\tr_\mathbb{R} \bP)\bI\, , 
\end{equation}
which we can also write, for non-zero mass, as 
\begin{equation}
\tilde\bP = \pm (\det\bP)\, \bV^\dagger \bV\, , 
\end{equation}
where $\bV$ is the inverse\footnote{The left and right inverse are equal, even for $\bK=\bH$.} to $\bU$:
\begin{equation}
\bV \bU = \bU\bV = \bI\, , \qquad \bV \to  \bN^\dagger \bV \bL^\dagger \, . 
\end{equation}

Using $\tilde\bP$ instead of $\bP$ we may construct the PL tensor
\begin{equation}
\bZ_- = \tilde\bP \bJ - \bJ^\dagger \tilde\bP = \mp (\det\bP) \bV^\dagger \bG \bV \, . 
\end{equation}
For $d=4$ this is just the co-vector version of the vector $\bZ_+$ but for $d=6$ it is an  anti-self-dual PL 3-form (assuming $\bZ_+$ to be self-dual).   The PL tensors $\bZ_\pm$ are related by
\begin{equation}
\bZ_+ \tilde\bP =  \bP \bZ_- \, , \qquad \tilde\bP \bZ_+ =  \bZ_- \bP\, . 
\end{equation}
For $d=6$ this is equivalent  to a relation found in \cite{Routh:2015ifa} using $SU^*(4)$ notation. 

Using the mass-shell constraint in the form (\ref{twistmass}) we have 
\begin{equation}
\tr_{\mathbb{R}} (\bZ_+\bZ_-) = - m^2 tr_{\mathbb{R}} \bG_0^2\, . 
\end{equation}
Whereas the left hand side is, by construction a Poincar\'e Casimir, the right hand side is  proportional to the quadratic Casimir of the rotation group.
This is to be expected from the fact that the Poincar\'e group representations are induced, for massive particles, by those of the rotation group.

\subsection{The 6D quartic Casimir}

For $d=6$ we still need to consider the PL  5-form $\Upsilon$ that is quadratic in $\bJ$; this is equivalent to a pseudo-vector that we shall call $y$
and represent by a quaternionic hermitian matrix  $\bY$.  As $p\cdot y=0$, an obvious guess is that 
\begin{equation}
\bY = \pm \bU \left[ \bG_0^2 - \frac{1}{2} \tr_{\mathbb{R}}(\bG_0^2)\right]\bU^\dagger\, ,  
\end{equation}
but we need to show that this expression can be rewritten as a polynomial in Poincar\'e Noether charges. Substitution for $\bG_0$ yields 
\begin{equation}\label{Ys}
2\bY= \left[\bY_ + -  \frac{\tr_{\mathbb{R}} (\bP\bZ_- \bJ)}{\det(\bU\bU^\dagger)}\,  \bP\right] + 
(\det\bP)^{-1} \bP \left[ \bY_- +  \frac{ \tr_{\mathbb{R}} (\bP\bZ_- \bJ)}{\det(\bU\bU^\dagger)} \tilde \bP\right]\bP\, , 
\end{equation}
where
\begin{equation}
\bY_+  = \bZ_+ \bJ^\dagger -\bJ\bZ_+ \, , \qquad  \bY_- = \bZ_- \bJ -\bJ^\dagger \bZ_- \, . 
\end{equation}

In the rest frame we have (supposing $m$ to be positive) that 
\begin{equation}
\bP= -\tilde\bP = \pm m \bI \qquad ({\rm rest \ frame})\, . 
\end{equation}
In this frame we have
\begin{equation}
\tr_{\mathbb{R}}\left[ \bY_- +  \frac{ \tr_{\mathbb{R}} (\bP\bZ_- \bJ)}{\det\bP} \tilde \bP\right] =0  \qquad ({\rm rest \ frame})\, . 
\end{equation}
This fact allows us to make use of the following lemma: 
 \begin{itemize}
 \item {\bf Lemma}: Given a hermitian matrix $\bM$ (over $\bK=\bR,\bC,\bH$) transforming as $\tilde \bP$ and such that $\tr_{\mathbb{R}}(\bM) =0$ in the rest-frame, then 
 \begin{equation}
 \bP \bM \bP = (\det\bP)\, \tilde \bM \, . 
 \end{equation}
 {\it Proof}: both sides transform as $\bP$ and are equal in the rest frame.
 \end{itemize}
Applying this lemma for $\bM$ equal to the matrix $[\bY_- + \cdots]$ appearing in (\ref{Ys}) we find that 
\begin{equation}
\bY = \frac{1}{2}\left[ \bY_+ + \tilde\bY_-\right]\, . 
\end{equation}
This expression is  polynomial in the Noether charges, so $\bY$ is indeed a PL vector, as shown in \cite{Routh:2015ifa} using $SU^*(4)$-spinor notation. 
We may use it to construct the quartic Poincar\'e Casimir
\begin{equation}
\tr_{\mathbb{R}}(\tilde \bY\bY) = m^2\tr_{\mathbb{R}} \left[\bG_0^2 - \frac{1}{2} \tr_{\mathbb{R}} \bG_0^2\right]^2\, , 
\end{equation}
where the mass-shell constraint is used to get the right hand side.

\subsection{From PL to SPL}

In the (super)twistor formalism of 3D,4D and 6D particle mechanics, the generalization from PL tensor  to SPL tensor is immediate.  We have now seen that one gets all 
PL tensors relevant for 3D, 4D and 6D by an appropriate ``dressing'' of powers of the spin-shell constraint matrix.  We arrived at this result by considering the bi-twistor action for a massive particle of zero spin,  with  spin-shell constraint matrix $\bG_0$.  In the context of superparticle mechanics we get the analogous SPL tensors in the same way from the bi-supertwistor action for the massive CBS superparticle, which has zero superspin.  We just have to interpret $\bG_0$ as the spin-shell constraint matrix of this superparticle action.

\section{Discussion}

Elementary particles are associated with irreducible unitary representations of the  Poincar\'e group, which  are classified by a mass $m$ and the 
eigenvalues of a set of Casimirs that  determine the spin.  In four-dimensional  Minkowski spacetime there is only one such spin Casimir and it is the norm  of the translation-invariant Pauli-Lubanski (PL) pseudo-vector, which is equivalent to a 3-form. For the super-Poincar\'e algebra,  the original construction by Salam and Strathdee \cite{Salam:1974yz} of the analogous superspin  Casimir  proceeded differently because of difficulties in constructing a supertranslation invariant extension of the  PL pseudo-vector.  

These difficulties were partially circumvented by Buchbinder and Kuzenko \cite{Buchbinder:1998qv} and by Pasqua and Zumino \cite{Pasqua:2004vq} via the proposal that a set of supertranslation invariant constraints should be imposed on the super-Poincar\'e charges. However, as one of these was $P^2=0$ the method was limited to massless particles. We have shown how to generalise the construction to non-zero mass by consideration of an implicit  BPS-saturated {\it extended} super-Poincar\'e algebra, and we have also explained how
the final results agree, where applicable,  with both  the Salam-Strathdee and Finkelstein-Villasante constructions.

We have also shown how the super-Pauli-Lubanski 3-form resulting from our construction arises naturally in the supertwistor formulation of superparticle mechanics. This is because the required constraints  on super-Poincar\'e Noether charges  become identities in this formulation, and the super-Pauli-Lubanski 3-form becomes a ``dressed'' version of the spin-shell constraint functions that appear  in the simplest (CBS) superparticle action; this is perhaps the simplest way to see why the quantum CBS superparticle has zero superspin. 

It also suggests that the supertwistor formalism is ideally suited for the determination of the superspin for generic superparticle mechanics models. As confirmation of this suggestion, 
we considered a simple ``superspinning particle'' action and used its supertwistor formulation to show that the quantum superspinning particle  has superspin $\tfrac{1}{2}$.  There is an obvious generalization to an ``extended superspinning particle'' modelled on the massive spinning particle with  $N>1$ local worldline supersymmetries \cite{Howe:1989vn}, for which a  supertwistor  formulation was given in \cite{Mezincescu:2015apa}. We expect this to have a superspin content that is the same as the spin content of its non-supersymmetric analog. 
 
As confirmed in section \ref{sec:three}, the supertwistor formulation of massive ${\cal N}=1$ superparticle models makes manifest a ``hidden''  ${\cal N}=2$ supersymmetry \cite{Mezincescu:2014zba}, which is the implicit BPS-saturated extended supersymmetry mentioned above.  Whether this additional supersymmetry survives quantization depends on whether a reality condition is imposed on the particle's wavefunction.  For example,  we found that quantization of the massive 
CBS superparticle  yields the ${\cal N}=1$ supermultiplet of superspin zero, which has helicity content $(-\tfrac{1}{2}, 0,0, \tfrac{1}{2})$,   because we implicitly chose to 
ignore the ``hidden'' supersymmetry.  If we had chosen to quantize preserving the BPS-saturated ${\cal N}=2$ supersymmetry then we would have found the 
${\cal N}=2$ hypermultiplet, which has the helicity content of a {\it doubled} ${\cal N}=1$ superspin-zero supermultiplet.  Imposing a reality condition eliminates this doubling 
and breaks the ${\cal N}=2$  supersymmetry to ${\cal N}=1$ supersymmetry. 

The same quantum option arises for the superspinning particle. Our claim that its quantization yields  the supermultiplet with superspin $\tfrac{1}{2}$ implicitly assumed a quantization preserving only the original ``built-in'' ${\cal N}=1$ supersymmetry. If instead we had quantized preserving ${\cal N}=2$ supersymmetry then we would have found a doubled  helicity content.  Precisely this doubled superspin-$\tfrac{1}{2}$ spectrum was found previously from quantisation of an apparently very different  ``massive spinning superparticle''  which has a ``built-in''  BPS saturated  ${\cal N}=2$  supersymmetry  \cite{KowalskiGlikman:1989eb,Bergshoeff:1989uj}.  This quantum coincidence suggests an equivalence between 
the  ``massive spinning superparticle'' and our  ``superspinning particle''.   In fact, this equivalence can be proved by adapting the proof in \cite{Mezincescu:2014zba} for the ``non-spinning'' case, which is based on a gauge-fixing that breaks ${\cal N}=2$ to ${\cal N}=1$ supersymmetry. 
 
 Our 4D results complement those obtained for the 6D massive superparticle in  \cite{Routh:2015ifa}, where the relation between spin-shell constraints and (super-)Pauli-Lubanski tensors was  also explored.  Here we have shown how this relation can be understood in a unified way for  Minkowski spacetimes of dimension $d=3,4,6$ by formulating the (super)particle in these dimensions in terms of $Sl(2;\bK)$ spinors, where $\bK=\bR,\bC,\bH$ are the three associative normed division algebras over the real numbers.  
 
 As the {\it massless} 10D superparticle can be written in $Sl(2;\bO)$ 
 spinor notation \cite{Oda:1988sz} it seems likely that there exists an $Sl(2;\bO)$ bi-spinor formulation of the massive 10D superparticle.  If so,  it  would be of interest if some of the 
 results reported here could be extended to 10D by means of an $Sp(4;\bO)$ twistor reformulation, but we leave this to future investigations.


\section*{Acknowledgements} We are grateful to Eric Bergshoeff for helpful correspondence and discussions. 
A.S.A. and P.K.T.  acknowledge support from the UK Science and Technology Facilities Council (grant ST/L000385/1), and A.S.A. also acknowledges support from Clare Hall College, Cambridge, and from the Cambridge trust. A.S.A. is grateful to Konstantinos Sfetsos for hospitality at the Faculty of Physics of the University of Athens during the revision of this paper, and L.M. gratefully acknowledges the hospitality of the Department of Applied Mathematics and Theoretical Physics of the University of Cambridge over the same period.


\providecommand{\href}[2]{#2}\begingroup\raggedright\endgroup

\end{document}